\documentclass{article}
\usepackage[utf8]{inputenc}
\usepackage{authblk}
\usepackage{hyperref}
\usepackage{bm}

\title {Retrospectively Diagnosing Einstein with Asperger's Syndrome and the Dismal Failure of Debunking Myths}
\author{Galina Weinstein}
\affil{\normalsize Reichman University, The Efi Arazi School of Computer Science, Herzliya; University of Haifa, The Department of Philosophy, Haifa, Israel.} 

\begin{document}

\maketitle

\begin{abstract}
In 2003, Simon Baron-Cohen, a world expert on autism, diagnosed Einstein posthumously with Asperger's syndrome. I think we cannot diagnose a dead person. Historians of science have fiercely objected to this trend of diagnosing deceased scientists by reconstructing from scant evidence, calling these diagnoses myths. Despite the historians' efforts at demolishing myths, Einstein has been constantly diagnosed with Asperger’s syndrome. I will stick my neck out and suggest in this paper that although historians' critique of Baron-Cohen and others includes debunking myths, it piggybacks on another myth that uses the following metaphors: a dull and socially adept Einstein who worked at Zurich, Prague, Berlin, and Princeton, an industrious scientist who earned his living through his work as a professor at the university; he had a special gift of friendship and collegiality, and he was deeply embedded in the academic community. These explanations do not make sense from the perspective of Einstein sitting in his office at Princeton, let alone Einstein sitting in the patent office. This perhaps explains the tendency of people to find counterclaims and myths more persuasive than historians' explanations which seem deeply problematic. 
\end{abstract}

\section{Introduction}

With some apprehension, I lectured about retrospectively diagnosing Einstein with autism (Asperger's syndrome) at the conference "Actually Autism" at Ben-Gurion University. I was reluctant to meddle in Einstein's private life. I prefer to occupy myself with Einstein's physics. I decided to get involved in this topic and give this lecture because in trying to debunk myths, people simply spread stigmas and prejudices about Asperger's syndrome and twist Einstein's biography. This paper presents an adapted version of my lecture. 
\vspace{2mm} 

Why does everybody ask if Albert Einstein had Asperger's syndrome? Because -- as I show in section \ref{2} -- in 2003, Simon Baron-Cohen made a posthumous diagnosis of Einstein, suggesting he showed many signs of Asperger's syndrome.\footnote{Here is a list of signs of twice-exceptionality and Asperger's syndrome in adults (including the following but not limited to these):
1. Lack of cognitive empathy (finding it hard to understand what others are thinking and feeling) and high empathic concern (desire to reduce other people's suffering and help them at your own expense). 
2. Have the same daily routine and get anxious if it changes. 
3. Have meltdowns (tantrums), difficulty managing emotions leading to outbursts, and may seem blunt and rude.  
4. Are autodidacts (can teach themselves just about anything). 
5. May have weak short-term memory and very strong long-term memory. May have a visual memory. 
6. Can be atheistic or leave one religion and enter another.
7. Have a spiky profile of strengths and weaknesses. 
8. May have dropped out of high school and returned later; may have unfinished or partial degrees. May have dropped out of classes and not attended classes.  
9. May have difficulty accepting criticism, stubbornness, and fixation on topics.
10. Having problems expressing themselves clearly, and talking and laughing very loudly, though there is little to laugh about.
11. Attention to details. 
12. Tends to burn bridges with people and may rebel against authority. 
13. Considered a "loner" and needs more time than others to be alone; has difficulty socializing for extended periods; wants to spend time socializing but is overstimulated by too many people. Does not just focuses around the clock on a special interest that may involve the person's career. Tends to be easily overwhelmed by noises and in social situations and then withdraws to a quiet spot; needs more time alone than others. 
14. Has a high sense of justice and fairness and is too honest. 
15. On the one hand, they can be witty and sarcastic and tend to be more childish; on the other, they can take things very literally and may not understand sarcasm, irony, or humor from other people. May use humor to overcome social obstacles by playing tricks on others. 
16. Have a remarkable power of concentration and can focus for long periods of time on the special interest without eating or drinking.
17. Unusual tolerance for unconventional and unorthodox thinking and people.
18. May dress differently than their peers; often appear eccentric; dress more for comfort than appearance. 
19. May invent limericks and send them to people they care about.
20. May feel things profoundly and experience intense emotions. 
From: Autism Awareness Centre Inc. } Baron-Cohen has argued that in several biographies of Einstein, it was written that Einstein was a loner and could not speak until he was three. Baron-Cohen did not base his retrospective diagnosis on historical medical records written by physicians. He instead diagnosed Einstein based on fragments of evidence mainly found in biographies. 
\vspace{2mm} 

As I discuss in section \ref{3}, several caveats exist in diagnosing a dead person retrospectively. The danger with historical retro-diagnosis is that the details of Einstein’s thinking process are not always precise, and we cannot extract a few details from a biography. Indeed, as I show in this paper, medical historians object to this practice saying that experts usually read a biography, extract some passages dealing with medicine, and come up with a diagnosis. What is even more insidious is that retro-diagnosis is made by people who are amateur historians rather than historiographical-trained scholars. They, therefore, even often draw on dubious sources to make historical diagnoses. Further, people retrospectively diagnosing historical figures are susceptible to biases. A controversy has erupted over the retro-diagnosis of Einstein. Historians of science have fiercely objected to this trend of diagnosing deceased scientists retrospectively by reconstructing from scant evidence. Despite the historians' warning, Einstein has been constantly diagnosed with Asperger’s syndrome. 
\vspace{2mm} 

In the end, in section \ref{1}, I show that there is a twist to the story. It turns out that historians are susceptible to biases, precisely like any other scholar. 

\section{Baron-Cohen diagnoses Einstein} \label{2}

It all started when Simon Baron-Cohen, a professor of developmental psychopathology and a world expert on autism at Cambridge University, suggested
that Einstein might have shown signs of Asperger’s syndrome \cite{Baron-Cohen1}. Baron-Cohen supports his thesis by 1) quoting an account by Einstein’s son Hans Albert written by Peter Michelmore in his biography of Einstein, \emph{Einstein, Profile of the Man},\footnote{Michelmore begins his book by saying: "In February, 1962, I spent two days with Hans Albert Einstein in his home overlooking San Francisco Bay. Hans Albert, fifty-seven, the older son of Albert Einstein [...]. He had never discussed his father before with any writer, at least not in depth. But he answered all my questions and waited while I wrote down the answers."} 
and 2) cherry-picking Einstein’s idiosyncratic habits from Roger Highfield and Paul Carter's book, \emph{The Private Lives of Albert Einstein}.\footnote{The authors base their stories on family members' reports and memories of people who knew Einstein.} Baron-Cohen writes \cite{Baron-Cohen1}:

\begin{quote}
Einstein was described as "lonely and dreamy" as a child, with difficulty in making friends. He was said to prefer "solitary and taxing" games, such as complex constructional play with blocks or making houses of cards up to fourteen stories high. He would "softly repeat every sentence he uttered—a habit he continued until the age of seven." He was still not considered fluent in speech at the age of nine. He was also a loner: "I'm not much with people," he would say. "I do not socialize because social encounters would distract me from my work and I really only live for that, and it would shorten even further my very limited lifespan."    
\end{quote}

This is almost verbatim what Highfield and Carter write in their book.\vspace{1mm} 
\footnote{The first part of Baron-Cohen's quotation is based on the following passage from Highfield and Carter \cite{Highfield}: 

\begin{quote}
By his description [to Bela Kornitzer], Einstein was a lonely and dreamy child who did not easily find companions. He would avoid the rough-and-tumble games when the children of relatives came to play in the Einsteins' garden, [...]. Besides assembling complicated constructions with his building blocks, he made houses of cards up to fourteen stories high. [...] After starting to talk comparatively late, he would softly repeat every sentence he uttered - a habit that continued until he was seven. Even when he was nine he lacked fluency of speech. The problem seems to have been as much a reluctance to communicate as any inability to do so.    
\end{quote}

The second part of Highfield and Carter's paragraph ("Besides assembling [...]") is based on Maja Winteler-Einstein's (Einstein's sister) biography of her brother \cite{Winteler-Einstein}. 

The second part of Baron-Cohen's quotation is based on the following passage from Highfield and Carter \cite{Highfield}: 

\begin{quote}
'I’m not much with people, and I’m not a family man.' he told Salaman. 'I want my peace. I want to know how God created this world.'    
\end{quote}

Esther Salaman was a young Jewish woman who interviewed Einstein in Berlin but published the interview thirty years after Einstein died in 1955. And the full quote is \cite{Salaman}: 

\begin{quote}
”I’m not much with people, and I’m not a family man. I want my peace. I want to know how God created this world. I am not interested in this or that phenomenon, in the spectrum of this or that element. I want to know His thoughts; the rest are details.”    
\end{quote}

The phrase "God created the world," so often repeated by Einstein, represents his religious feeling that laws of nature could be formulated in the simplest form.
\vspace{1mm} 

The quotation "I do not socialize because social encounters would distract me from my work" is taken from the following source. Einstein’s physician and friend Janos Plesch writes \cite{Plesch}: 

\begin{quote}
On suitable occasions, one could also get from him [Einstein] an uninhibited opinion about his immediate circle. During one such open phase during our conversation, I told him his environment in America and asked, 'Who are your friends in this small town?' He replied 

\begin{quote}
[…] I do not socialize because social encounters would distract me from my work and I really only live for that, and it would shorten even further my very limited lifespan. I do not have any close friends here as I had in my youth or later in Berlin with whom I could talk and unburden myself. That may be due to my age. I often have the feeling as if God has forgotten me here.    
\end{quote}
\end{quote}

} 

\noindent In his lecture “Scientific talent and autism: is there a connection?”, Baron-Cohen merely repeats the arguments put forward in his previous text \footnote{In the case of Einstein, says Baron-Cohen,  
\begin{quote}
"if you look at his childhood, he is described as being alone, having no friends, and not being interested in mixing and socializing. Apparently, he was late to talk, a language delay. Apocryphally he didn’t speak until he was five years old. And even then, when he did start to speak, his speech showed echolalia. And as an adult and as a distinguished scientist, he said: ‘I do not socialize because that would distract me from my work, and I really only live for that.’ 
\end{quote}}
and adds \cite{Baron-Cohen2}:
\begin{quote}
So he wanted to be away from people and to only focus (some people would say obsessionally) on the world of physics, and he also had a strong interest in music when he wasn’t doing physics. And he said: ‘Music is a way for me to be independent of people.’ So you can see that this was a man who didn’t want to spend time socializing and instead wanted to focus on physics and music. 
\end{quote}

Again, Baron-Cohen cherry-picks quotations from Highfield and Carter \cite{Highfield} and Michelmore \cite{Michelmore}.\footnote{The quotation: "[...] music is a way for me to be independent of people" was taken from Michelmore \cite{Michelmore}:

\begin{quote}
Einstein worked alone now, so music was the only way to get close to him. That was how Manfred Clynes "reached" Einstein. [...] Helen Dukas called to invite me [Cylnes] to supper. [...] Einstein himself was rather jovial. Over the meal of meat and vegetables and for two hours afterward, we talked about music. He told me it was as important for him to improvise on the piano as it was for him to work on his physics. 'It is a way for me to be independent of people,' he said. 'And this is highly necessary for the kind of society in which we have.' 
\end{quote}}

\vspace{3mm} 

Barbara Wolff -- who had a long tenure at the Albert Einstein Archives and is an editor in the Einstein Papers Project -- and Hananya Goodman write \cite{Wolff}:

\begin{quote}
Some of the characterizations of AS [Asperger's syndrome] described in the paragraph above actually apply well to the young Albert as we know him from Maja’s and Max Talmey’s recollections.

Both Maja and Talmey describe a boy who took little interest in boisterous games and, in general, in his peers, a boy who would concentrate patiently on elaborate constructions with building blocks or playing cards, delve into books and tricky arithmetic problems or play the violin. A sort of glass pane, as he called it many years later, separated him from his fellow human beings. Had such "social phobia" then been classified as a personality disorder, and had his parents and doctors felt the need to ‘heal’ the boy by making him conform to some norm, Albert might not have become Einstein.

\end{quote}

It seems that Wolff and Goodman agree with Baron-Cohen because, according to the latter, Albert has become Einstein because he is an Asperger. 
 
\vspace{3mm} 

\section{Historians object to diagnosing Einstein with autism} \label{3}

I think we cannot diagnose a dead person. Historians of science and medicine have fiercely objected to Baron-Cohen's retrospective diagnosis and have generally leveled criticism at clinicians, psychologists, and non-experts for retrospectively diagnosing disorders and diseases in deceased scientists. Below I provide the major objections made by historians:
\vspace{3mm} 

1. Historians argue that people sometimes draw on dubious sources to make retrospective diagnoses \cite{Kean}. For instance, Daniela Caruso, professor of law from Boston University, diagnosed Einstein with Asperger's syndrome based on a children's book. She writes \cite{Caruso}:

\begin{quote}
Post-mortem diagnoses are doubtful, but Albert Einstein’s life story, which begins with tales of delayed speech and abysmal performance at school, suggests that the most accomplished scientist of all time might have suffered from Asperger-like symptoms.     
\end{quote}

Caruso refers in a footnote to: “Don Brown, \emph{Odd Boy Out: Young Albert Einstein} (2004)” \cite{Caruso}. 
\emph{Odd Boy Out: Young Albert Einstein} is a children's book by Don Brown. 
\vspace{1mm} 

\noindent I am disappointed with the cheapening of history. 

\vspace{3mm} 

2. The retrospective diagnosis is a hobby of clinicians interested in history who enjoy testing their diagnostic acumen on famous historical figures. The physicians base their diagnosis on scant evidence, usually described in a non-medical context. They uncover intriguing medical secrets in reports that belong to an accessory section for incidental topics or letters to the editor in medical journals.  
This practice raises ethical issues; in particular, the dead scientists cannot say no. Retro-diagnosis can expose personal details that no one in Einstein's time could have known or understood. The retrospective diagnoses could pose ethical problems or questions that, if revealed, may cause harm to the scientist's living descendants and relatives. 
The diagnosing physicians strongly violate the principles of the medical profession because they give their retrospective opinion on a patient they have never seen nor examined \cite{Muramoto}, \cite{Kean}, \cite{Karenberg}. 

Baron-Cohen replied to the above criticism in his lecture “Scientific talent and autism: is there a connection?” \cite{Baron-Cohen2}:

\begin{quote}
This biographical approach is very interesting, but [...] I’ve suggested that this might be a very unreliable way to approach the question because biographies are always fragmented. We only have partial information about what the person was like, information that has survived historically. The person isn’t here to speak up for themselves, so we don’t really know whether, if we saw them today alive, they would meet the criteria of autism or Asperger’s syndrome. But certainly, it's pointing at the connection between great scientific talents and autism or Asperger’s”.    
\end{quote}    

There are two problems here: the first problem is the impossibility of verifying or falsifying the retrospective diagnosis for obvious reasons: we cannot examine and test the historical subject. And the second problem is ethical: publishing a diagnosis of a patient with whom the medical expert never had a physician-patient relationship and without consent. 

One could argue that historical celebrities are immune from privacy protection because their lives are open to the public. And because they are long dead, the subject of any potential harm no longer exists. But certain defenders of the celebrity's image and descendants of the historical figure might object to the publication of a diagnosis that potentially stains the reputation of the historical figure \cite{Muramoto}.

John Stachel gave a name to people who care about Einstein’s legacy (what he considered as the staunchest defenders of his reputation); he called them "keepers of the flame".\footnote{Stachel writes in his book, \emph{Einstein From 'B' to 'Z'}: "I soon became aware of another peril involving loss of boundaries: the danger of becoming a ‘keeper of the flame’ rather than a seeker of the truth." \cite{Stachel}}  

As a result of criticism, Baron-Cohen admitted there was a problem diagnosing a person after his death because one needs the person's consent. In his 2020 book, \emph{The Pattern Seekers: A New Theory of Human Invention}, Baron-Cohen attenuates his argument \cite{Baron-Cohen3}:

\begin{quote}
“Some hyper-systemizers,\footnote{Hyper-systemizing is part of the cognitive style of autistic people. \cite{Baron-Cohen4}} in a range of fields, have been described [by the mathematician Ioan Mackenzie James from Oxford University] as autistic. For example, [...] Albert Einstein and Henry Cavendish in the field of physics have all been described [by James] as autistic. In my view, it is unhelpful to speculate if someone – living or not – might be autistic, since a diagnosis is only useful if the person is seeking help and is struggling to function. Diagnosing someone – living or not – on the basis of fragmentary biographical information is unreliable and arguably unethical since diagnosis should always include the consent of the person and be initiated by them.
And from a scientific perspective, hyper-systemizing does not automatically mean you’re autistic”.     
\end{quote}

Indeed, on January 17, 1927, Hugo Freund wrote to Einstein to ask if he would allow himself to be psychoanalyzed (\cite{CPAE15}, Doc. 457). It is not known if an answer was sent, but on the letter, in German in Einstein’s handwriting, there is the following draft of a reply \cite{CPAE15}, Doc. 458; \cite{Dukas}: 
“I am sorry not to be able to comply with your request, as I would very much like to remain in the darkness of not having been analyzed.”     
In other words, Einstein refused to be diagnosed and psychologically analyzed. Notwithstanding that, Baron-Cohen and others have retrospectively diagnosed Einstein. 

But in the above text, Baron-Cohen tries to shrug off the diagnosis of Einstein with Asperger's syndrome, saying that James described Einstein and others as autistic. In 2012, Baron-Cohen went so far as to publish a paper with several Cambridge University students having Asperger's syndrome and write: “Einstein, Mozart, Newton, Wittgenstein, and others (predominately male, white, and deceased) are identified posthumously with AS and brilliance” \cite{Baron-Cohen5}. 
This sentence is disturbing because it insinuates that the brilliant people are predominantly white males. Baron-Cohen and his students intertwine two diagnoses - Asperger's syndrome and brilliance - into white males.  

\vspace{3mm} 

3. Let us move on to the third argument. Historical diseases and conditions must be interpreted in their historical context. Retrospective diagnosis is anachronistic because people try to diagnose a disease or disorder of the past in contemporary terms. This is called "anachronistic diagnosis." 
Critics of clinicians making retro-diagnosis say this kind of diagnosis can be misleading. 

First, medical experts arrive at a diagnosis of a historical figure using modern diagnostic criteria; they reason that if historical figures had all these symptoms, then they could have had these specific conditions. But in the past, people had different habits and lifestyles than we do \cite{Muramoto}. Hence, we should investigate social diagnosis, i.e., the name that people in the past allocated to a condition \cite{Cunningham}. If the social circumstances allowing us to perceive autism did not exist before some point relatively early in the twentieth century, then to what extent can we say the condition existed previously? \cite{McDonagh} 

Second, medical knowledge itself changes over time. Writing historical papers on famous patients is problematic because clinicians regard the level of scientific knowledge at the time of writing but assess the course of an illness in the past.\footnote{For instance, the historical figure, Polish composer Frédéric Chopin was first (1899) diagnosed as having tuberculosis; then, in 1961, allergic conditions and valvular stenosis were offered as a better explanation for his condition. In 1987 he was diagnosed with cystic fibrosis; in 1994, other genetic defects were suggested, such as alpha-1-antitrypsin-deficiency \cite{Karenberg}.} 

\vspace{1mm} 

For example, Temple Grandin is diagnosed with autism and has a Ph.D. in animal science. She concludes, “Einstein had many traits of an adult with mild autism, or Asperger’s syndrome.” And \cite{Grandin}:
\begin{quote}
As a child, Einstein had many of these traits. He did not learn to speak until he was three. In a letter to a mother of an autistic child, he admitted to not being able to learn to speak until late and that his parents had been worried about it.    
\end{quote}

This passage is anachronistic. How does Grandin know that both the child and Einstein were autistic? Children were not yet diagnosed as autistic in Einstein's lifetime.  
\vspace{3mm} 

4. Historians object to a retrospective diagnosis that does not provide new medical knowledge. This is as opposed to research based on historical sources that will help modern medical researchers understand the spread of diseases around our planet during human evolution and help us plan for unexpected health events. There are many good reasons to study diseases and disorders in the past. For instance, epidemiological analysis of data from historical smallpox outbreaks has helped appropriate authorities plan for possible outbreaks in the future \cite{Mitchell}.

To this, clinicians have responded by saying that studying historical figures might help to determine how prevalent autism was in previous generations \cite{Altschuler}. But then historians try to rebut this claim - see objection 3. 

Concerning autism in previous generations, let us return to Temple Grandin. Grandin compares herself to Einstein: "Like Einstein, I am motivated by the search for intellectual truth." She then further elucidates on this point \cite{Grandin}:

\begin{quote}
When he developed the theory of relativity, he imagined himself on a beam of light. His visual images were vaguer than mine, and he could decode them into mathematical formulas. My visual images are extremely vivid, but I am unable to make the connection with mathematical symbols. Einstein’s calculation abilities were not phenomenal. He often made mistakes and was slow; but his genius lay in being able to connect visual and mathematical thinking. Einstein’s dress and hair were typical of an adult with autistic tendencies, most of whom have little regard for social niceties and rank. When he worked at the Swiss patent office, he sometimes wore green slippers with flowers on them”.    
\end{quote}

The above passage abounds in errors. But I nevertheless want to raise two comments on this passage: 
\vspace{2mm} 

First, the people purporting to compare themselves to Einstein fail to mention that his disheveled exterior reflected his inner humility. As he once said:
"I speak to everyone in the same way, whether he is the garbage man or the university president" \cite{Jerome}. This is typical of people with Asperger's syndrome, but by the same token, this behavior is also typical of humble people.  
Hence, I suggest \emph{not} comparing yourself to Einstein if only for the many things you don't know about him. You can't know everything about Einstein because it is impossible to read \emph{all} the primary sources. And even if you manage to read all that, many documents have gone missing, and there is always hope of finding new lost letters, diaries, and manuscripts.

I will give you an example. Einstein once told Salaman: “Very few women are creative. I should not have sent a daughter of mine to study physics. I'm glad my wife doesn't know any science, although my first wife did." 
Grandin (a female scholar), do you still want to compare yourself to Einstein? This might be a slip of the tongue because Einstein admired several women and female physicists and mathematicians of his time, e.g., the brilliant genius physicist Marie Curie and genius mathematician Emmy Noether. 
Over the years, Einstein made different statements that depended on the people he spoke with. He would say one thing in the presence of distinguished scientists, cooperating in tributes without questioning the narratives, and would say quite the opposite on other occasions. Thus, I would be extremely careful when comparing someone to Einstein.
\vspace{2mm} 

Second, after Einstein's rise to fame, Max Flückiger wrote in his book \emph{Einstein in Bern} that Einstein often appeared in green slippers trimmed with flowers at work. The patent office officials called him "The man with the green slippers." A colleague also recalled that Einstein showed up at the patent office one day with a saw and proceeded to shorten the legs of his chair because it was not adjustable and was too high for him \cite{Fluckiger}. Abraham Pais described Flückiger's book as containing several reproductions of rare documents about Einstein's younger days. But "The text contains numerous inaccuracies" \cite{Pais}. In other words, Flückiger concocted the above stories out of things that never happened.
\vspace{1mm} 

5. Historians argue that it is often unclear how a highly specific diagnosis would make any difference in the scholarship of the historical figure in question \cite{Muramoto}.
It is asked: what is the objective of retrospectively diagnosing Einstein? What goals does it serve? What justifies the retrospective diagnosis of historical figures? What is the reason modern researchers retrospectively diagnose a historical figure as autistic? Does the diagnosis of Einstein as autistic deliver new medical insight into autism? Does it help us understand how Einstein created the theory of relativity and how he arrived at his significant discoveries?  

Ioan James writes \cite{James3}: "In the case of Einstein, we can conclude that he did have Asperger’s syndrome." James has further stated: "Although it seems to be widely accepted that Einstein had the syndrome, none of the many detailed biographies mentions this” \cite{James}. First, such a diagnosis does not make any difference in the scholarship of Einstein and the pathway to his theories. 
Second, not only do none of the "biographies mention this," several biographies even try to denounce it, as I show in the next section, claiming it is an utter myth.

\section{Historians try to debunk myths about Einstein} \label{1}

It seems to me that in trying to debunk myths, some historians hobble the narrative of extraordinary and creative scientists in that they extend the life of other myths. 
The issue is that it is not so much that historians endeavor to debunk myths that bothers me; it is how they do so.

Although historians’ critique of Baron-Cohen and others includes debunking myths, it piggybacks on another myth: presenting Einstein as a dull, conscientious, agreeable person and an industrious and socially adept individual better suited to Thomas Kuhn's incremental "normal science."
\vspace{3mm} 

First, when I wrote my first book, I was told that Albert Einstein, his sister Maja Winteler-Einstein, and the family members had exaggerated the story of Albert, who developed slowly, learned to talk late, and whose parents thought something was wrong with him. I was further told that, of course, there is no doubt that these stories have a grain of truth, and Maja and Albert recount their recollections in all sincerity, but these stories sound like family tales and may be exaggerated. Moreover, Einstein’s friends and assistants (in interviews and correspondences) contributed to spreading this myth. They thus inspired biographers to create a widespread mythical public image of Albert Einstein that embodies stories about Einstein, "the retarded genius."

We first conjecture that Einstein and his family members had exaggerated the story of Albert, "the retarded genius," and then show that the "real" Albert was a different matter. For instance, Albert's mother recognized that her son was talented, and it does not seem that she thought he would develop into some eccentric.    
But seen from a distance of time, almost ten years after writing my book \cite{Weinstein}, the above argument sounds improbable because Einstein and his family are accused of inventing stories to create a myth that serves a specific purpose. 
\vspace{3mm} 

A different way of considering the situation is presenting Einstein as having a special gift of friendship. He may not have been working with a large group of scientists. Still, he was deeply embedded in the academic community and was a modern professional scientist who earned his living through his work as a professor at the university. Historians debunk myths by saying that Einstein was not an isolated genius "working by himself in an attic with pen and paper" \cite{Fesenmaier}. Einstein had close friendships with physicists and a special gift of collegiality and was very well-adjusted. 
Then the argument goes on like this. Although Einstein was sometimes rude or insensitive, he was not unlike many individuals. Some of his social behaviors and attitudes may seem unconventional today, but they were not unusual in his day and milieu. He was very sociable and was emotionally involved with friends and family, even if he savored solitude when it was within his reach to pursue his science \cite{CKS}.\footnote{I would like to correct some of the stigmas and prejudices about Asperger's syndrome found in \emph{An Einstein Encyclopedia}. Asperger's syndrome (autism) is neither a disease nor a mental or emotional disorder. Asperger's syndrome (autism) is also not a disorder people "suffer" from. Asperger's syndrome is \emph{not} any of the following: a defect, developmental anomaly, pathological conduct, a mental illness, schizophrenia, and bouts of severe depression.
} 

In his biography of Einstein, \emph{His Life and Universe}, Walter Issaacson elucidates why he is not convinced of the diagnosis of Einstein with Asperger's syndrome: "Even as a teenager, Einstein made close friends, had passionate relationships, enjoyed collegial discussions, communicated well verbally, and could empathize with friends and humanity in general" \cite{Issacson}. \footnote{Issacson writes: "A Google search of Einstein + Asperger's results in $146,000$ pages. I do not find such a long distance diagnosis to be convincing" \cite{Issacson}. I did the exact search, and I found only $25$ pages. Still, it does not mean that these $25$ pages are convincing.} 

In my opinion, the above narrative, very unfortunately, strips off Einstein from his great sense of humor, from being stubborn and unconventional and sticking out his tongue at his pursuers to express his annoyance (the photograph that has been reproduced endlessly). 

Einstein's former Cantonal School classmate, Hans Byland, one year older than Einstein, had painted a verbal portrait of Einstein as a teenager: Einstein as a young man could not be fitted into any pattern, an impudent Swabian, a restless spirit; nothing escaped the sharp gaze of his large bright brown eyes. "A sarcastic curl of his rather full mouth with the protruding lower lip did not encourage Philistines to fraternize with him." His attitude towards the world was that of a laughing philosopher, and his witty mockery pitilessly lashed out at any conceit or pose. Einstein made no bones about voicing his personal opinions, whether offensive or not, and he had courageous adherence to the truth \cite{Seelig}. Of course, one can call this "communicated well verbally." 
We have Einstein's rude letter in which he lashed out at the editor of \emph{The Physical Review}: "We (Mr. Rosen and I) had sent you our manuscript for publication and had not authorized you to show it to specialists before it was printed. I see no reason to address your anonymous expert's – in any case, erroneous – comments. Based on this incident, I prefer to publish the paper elsewhere" \cite{Kennefick}. And Einstein attacked his opponents at the Hebrew University and termed Judah Leon Magnes and his supporters “downright vermin.” On another occasion, he called the university a “bug-infested house” \cite{Rosenkranz}. It, therefore, seems that Einstein made no bones about voicing his opinions, whether offensive or not.  
\vspace{3mm} 

It seems to me that the supporters of the view of Einstein as a very well-adjusted individual not only fail in demolishing myths, this narrative repudiates Einstein's patent office years. 
Einstein's friends and colleagues at the patent office, his table and drawer, pen and sheets of paper, and especially the thoughts he hatched there were antipodal to an academic community. 

Einstein later wrote that, unlike himself, Marcel Grossmann "was not a vagabond and loner"\cite{Einstein}. Einstein wrote to Mileva Marić: “I always find that I am in the best company when I am alone, except when I am with you” \cite{CPAE1} (Einstein to Marić, Doc 128, Dec 17, 1901). Indeed, unlike Einstein, Grossmann was quickly given an assistant post under Prof. Wilhelm Fiedler; Grossmann succeeded Fiedler as a professor at the Polytechnic in 1907.     
On the other hand, three years after the publication of the miraculous year papers -- special relativity, $E = mc^2$ first paper, quanta and Brownian motion -- Jakob Laub wrote to Einstein, "I must confess to you that I was surprised to read that you have to sit in an office for eight hours a day. But, history is full of bad jokes" (Laub to Einstein, \cite{CPAE5}, Doc. 91, March 1, 1908). 

Suppose Einstein had this unique quality of being well-adjusted and deeply embedded in the academic community, as purported. Why did he sit in the patent office for so long? None of this makes sense in the picture of the well-adjusted and embedded-in-the-science-community Einstein. A narrative of a socially adept Einstein does not even make sense in the context of the later Einstein, sitting in his office in Princeton, let alone sitting in the patent office. 

We are trying to press a conscientious and agreeable Einstein into a rebellious Einstein. But we don't need Karl Popper to tell us that Einstein was one of the most exciting and imaginative scientists ever to live; he was intelligent and creative, a vocational revolutionary scientist. And this is of utmost importance. He was brilliant, inspirational, abrasive, and rebellious, precisely the type of truth-seeker to create a scientific revolution. \footnote{See the discussion in \cite{Charlton}.}
\vspace{3mm} 

Einstein thought that certain personal matters of his life should not be made public. That is the reason he did not like biographies that dealt too much with his personal life \cite{Weinstein}. But private letters are profusely found in the \emph{Collected Papers of Albert Einstein} (\emph{CPAE}); for instance, Einstein's love letters to his first wife, Mileva Marić; Einstein's later correspondence with Marić and his two sons, Hans Albert, and Eduard; Einstein's exchange of letters with his second wife and first cousin, Elsa Lowenthal Einstein; correspondence with Einstein’s sister, Maja; letters Einstein wrote to his extra-marital lover Betty Neumann, etc. The reasons behind the decision to publish Einstein’s private correspondence are that Einstein is not a private person anymore; his life is open to the public. Einstein has ceased being a private individual and belongs to history. Since Einstein is long dead, publishing his letters cannot cause him any harm. On the contrary, he is a historical scientist in the same category as Newton and Galileo. We do not think of the privacy of such historical figures. We want to know as much as we can about them because, without full knowledge, one cannot fully understand the evolution of their ideas \cite{Bailey}. 

But then, based on Einstein’s private correspondence, biography of his sister Maja (an abridged version of which was published in the \emph{CPAE}), and other first-hand evidence, people have written biographies and books on Einstein's private life. And based on these books, other people, in turn, are retrospectively diagnosing Einstein with all sorts of things. Historians who care about Einstein’s legacy then fiercely object to this abuse of Einstein’s privacy and, in turn, engage in (futile) efforts to protect his privacy. Hence, we have vicious circularity.  

\section*{Acknowledgement}

This work is supported by ERC advanced grant number 834735.


\begin{thebibliography}{38}

\bibitem[1] {Altschuler} E. L. Altschuler (2013). "Asperger’s in the Holmes Family." \emph{Journal of Autism and Developmental Disorders} 43, pp. 2238–2239.  

\bibitem[2] {Bailey} A. S. Bailey (1989). "On the Collected Papers of Albert Einstein: The Development of the Project." \emph{Proceedings of the American Philosophical Society} 133, pp. 347-359. 

\bibitem[3] {Baron-Cohen1} S. Baron-Cohen (2003). \emph{Male and Female Brains and the Truth about Autism. The Essential Difference}. New York: Basic Books.

\bibitem[4] {Baron-Cohen2} S. Baron-Cohen (2011). “Scientific Talent and Autism.” Lecture.

\bibitem[5] {Baron-Cohen3} S. Baron-Cohen (2020) \emph{The Pattern Seekers: How Autism Drives Human Invention}. New York: Basic Books.

\bibitem[6] {Baron-Cohen4} S. Baron-Cohen, E. Ashwin, C. Ashwin, T. Tavassoli, and B. Chakrabarti (2009). “Talent in autism: hyper-systemizing, hyper-attention to detail and sensory hypersensitivity.” \emph{Philosophical Transactions of the Royal Society B: Biological Sciences} 364, pp. 1377–1383.

\bibitem[7] {CKS} A. Calaprice, D. Kennefick, and R. Schulmann (2016). \emph{An Einstein Encyclopedia}. Princeton: Princeton University Press. 

\bibitem[8] {Caruso} D. Caruso (2010). “Autism in the U.S.: Social Movement and the Legal Change.” \emph{American Journal of Law and Medicine} 36, pp. 483-539.

\bibitem[9] {Charlton} B. G. Charlton (2009). "Why are modern scientists so dull? How science selects for perseverance and sociability at the expense of intelligence and creativity." \emph{Medical Hypothesis} 72, pp. 237-243. 

\bibitem[10] {CPAE1} \emph{Collected Papers of Albert Einstein. Volume 1: The Early Years, 1879–1902}. J. Stachel, D. C. Cassidy, and R. Schulmann (eds.). Princeton: Princeton University Press, 1987. 

\bibitem[11] {CPAE5} \emph{Collected Papers of Albert Einstein. Volume 5: The Swiss Years: Correspondence, 1902–1914}. M. J. Klein, A. J. Kox, and R. Schulmann (eds.). Princeton: Princeton University Press, 1993.

\bibitem[12] {CPAE15} \emph{Collected papers of Albert Einstein. Volume 15: The Berlin Years: Writings and Correspondence, June 1925-May 1927}. D. Kormos Buchwald, J. Illy, A. J. Kox, D. Lehmkuhl, Z. Rosenkranz and J. Nollar James (eds.). Princeton: Princeton University Press, 2018.

\bibitem[13] {Cunningham} A. Cunningham (2002). "Identifying disease in the past: cutting the Gordian knot." \emph{Asclepio} 54, pp. 13–34.

\bibitem[14] {Dukas} H. Dukas and B. Hoffmann (1979). \emph{The Human Side. New Glimpses from his Archives}. Princeton: Princeton University Press.

\bibitem[15] {Einstein} A. Einstein (1955). "Autobiographische Skizze." In C. Seelig (1956). \emph{Helle Zeit – Dunkle Zeit. In memoriam Albert Einstein}. Zurich: Braunschweig: Friedrich Vieweg und Sohn/Europa. 

\bibitem[16] {Fesenmaier} K. Fesenmaier (2014)."Einstein Online: An Interview with Diana Kormos-Buchwald," \emph{Caltech}.

\bibitem[17] {Fluckiger}  M. Flückiger (1974). \emph{Albert Einstein in Bern}. Switzerland: Verlag Paul Haupt Bern. 

\bibitem[18] {Grandin} T. Grandin (1995). \emph{Thinking in Pictures: My Life with Autism}. New York: Vintage Books.

\bibitem[19] {Baron-Cohen5} J. Hastwell, N. Martin, S. Baron-Cohen, and J. Hardling (2012). “Giving Cambridge University students with Asperger syndrome a voice: a qualitative, interview-based study towards developing a model of best practice.” \emph{Good Autism Practice} 13, pp. 56-63.

\bibitem[20] {Highfield} R. Highfield and P. Carter (1994). \emph{The Private Lives of Albert Einstein}. New York: St. Martin's Press.

\bibitem[21] {Issacson} W. Issacson (2007). \emph{His Life and Universe}. New York: Simon $\&$ Schuster.

\bibitem[22] {James} I. M. James (2003). “Singular scientists." \emph{Journal of the Royal Society of Medicine} 96, pp. 36–39.

\bibitem[23] {James3} I. M. James (2006). \emph{Asperger's Syndrome and High Achievement: Some Very Remarkable People}. London: Jessica Kingsley Publishers.

\bibitem[24] {Jerome} F. Jerome and R. Taylor (2006). \emph{Einstein on Race and Racism}. New Jersey: Rutgers University Press. 

\bibitem[25] {Karenberg} A. Karenberg (2009). "Retrospective Diagnosis: Use
and Abuse in Medical Historiography." \emph{Prague Medical Report} 110, pp. 140–145.

\bibitem[26] {Kean} S. Kean (2012). "Retrodiagnoses. Investigating the ills of long-dead celebrities". \emph{Science} 337, pp. 30–31.

\bibitem[27] {Kennefick} D. Kennefick (2005). "Einstein versus the Physical Review." \emph{Physics Today} 58, pp. 43-48.  

\bibitem[28] {McDonagh} P. McDonagh (2008). “Autism and Modernism. A Genealogical Exploration.” In M. Osteen (ed.). \emph{Autism and Representation}, pp. 99-116.

\bibitem[29] {Michelmore} P. Michelmore (1962). \emph{Einstein, Profile of the Man}. New York: Dodd, Mead $\&$ Company.  

\bibitem[30] {Mitchell} P. D. Mitchell (2011). "Retrospective diagnosis and the use of historical texts for investigating disease in the past." \emph{International Journal of Paleopathology} 1, pp. 81–88.

\bibitem[31] {Muramoto} O. Muramoto (2014). "Retrospective diagnosis of a famous historical figure: ontological, epistemic, and ethical 
considerations." \emph{Philosophy, Ethics, and Humanities in Medicine} 9, pp. 1-15.

\bibitem[32] {Pais} A. Pais (1982). \emph{Subtle is the Lord. The Science and Life of Albert Einstein}. Oxford: Oxford University Press.
 
\bibitem[33] {Plesch} J. Plesch and P. H. Plesch (1955). “Some Reminiscences of Albert Einstein.” \emph{Notes and Records of the Royal Society of London} 49, pp. 303-328.

\bibitem[34] {Rosenkranz} Z. Rosenkranz (2011). \emph{Einstein Before Israel: Zionist Icon or Iconoclast?} Princeton: Princeton University Press.

\bibitem[35] {Salaman} E. Salaman (1955) “A Talk With Einstein.” \emph{The Listener} 54, pp. 370-371.

\bibitem[36] {Seelig} C. Seelig (1956). \emph{Albert Einstein: A Documentary Biography}. Translated to English by Mervyn Savill. London: Staples Press. (1954). \emph{Albert Einstein. Eine dokumentarische Biographie}. Zurich: Europa Verlag.

\bibitem[37] {Stachel} J. Stachel (2002). \emph{Einstein From 'B' to 'Z'}. Boston: Birkhäuser.

\bibitem[38] {Weinstein} G. Weinstein (2015). {General Relativity Conflict and Rivalries: Einstein's Polemics with Physicists} (Newcastle, UK: Cambridge Scholars Publishing).

\bibitem[39] {Winteler-Einstein} M. Winteler-Einstein (1924). \emph{Albert Einstein – A Biographical Sketch}. \emph{Beitrag für sein Lebensbild}.\emph{Collected Papers of Albert Einstein. Vol. 1: The Early Years,1879–1902}. J. Stachel, D. Cassidy, and R. Schulmann (eds.). Princeton: Princeton University Press, 1987, pp. xv-xxii. (pp. xlviii-lxvi.) 

\bibitem[40] {Wolff}  B. Wolff and H. Goodman (2001). “The Legend of the Dull-Witted Child Who Grew Up to Be a Genius.” \emph{The Albert Einstein Archives} The Hebrew University of Jerusalem.

\end{thebibliography}
\end{document}